\documentclass[prd,showpacs,superscriptaddress,floatfix,nofootinbib,10pt]{revtex4-2}
\usepackage{times}
\usepackage{amssymb,amsbsy,amsmath,amsfonts}
\usepackage{graphicx}
\usepackage{float}
\usepackage{color}
\usepackage{orcidlink}
\usepackage{morefloats}
\usepackage{rotating}
\usepackage{srcltx}
\usepackage{slashed}
\usepackage{multirow}
\usepackage{verbatim}
\usepackage{hyperref}
\usepackage{tabularx}
\usepackage{bm}
\allowdisplaybreaks[4]

\usepackage{bbding}
\usepackage{threeparttable}

\usepackage{mathrsfs}

\newcommand{\PreserveBackslash}[1]{\let\temp=\\#1\let\\=\temp}
\newcolumntype{C}[1]{>{\PreserveBackslash\centering}p{#1}}
\newcolumntype{R}[1]{>{\PreserveBackslash\raggedleft}p{#1}}
\newcolumntype{L}[1]{>{\PreserveBackslash\raggedright}p{#1}}

\begin{document}

\title{The decay $\eta_c\to p\bar p\eta$ with meson--baryon final-state interactions}

\author{Jing Song\orcidlink{0000-0003-3789-7504}}
\email[]{Song-Jing@buaa.edu.cn}
\affiliation{Center for Theoretical Physics, School of Physics and Optoelectronic Engineering, Hainan University, Haikou 570228, China}
\affiliation{School of Physics, Beihang University, Beijing, 102206, China}
\affiliation{Departamento de Física Teórica and IFIC, Centro Mixto Universidad de Valencia-CSIC Institutos de Investigación de Paterna, 46071 Valencia, Spain}

\author{Yu-Shan Ren\orcidlink{https://orcid.org/0009-0001-4911-3829}}
\email[]{Yushan.Ren@ific.uv.es}
\affiliation{School of Physics, University of Electronic Science and Technology of China, \\Chengdu 610054, China}
\affiliation{Departamento de Física Teórica and IFIC, Centro Mixto Universidad de Valencia-CSIC Institutos de Investigación de Paterna, 46071 Valencia, Spain}
\vspace{0.5cm}

\author{Eulogio Oset\orcidlink{0000-0002-4462-7919}}%
\email[]{oset@ific.uv.es}
\affiliation{Departamento de Física Teórica and IFIC, Centro Mixto Universidad de Valencia-CSIC Institutos de Investigación de Paterna, 46071 Valencia, Spain}	
\affiliation{Department of Physics, Guangxi Normal University, Guilin 541004, China}

\begin{abstract}
We study the decay \(\eta_c\to p\bar{p}\eta\) by combining the SU(3)
flavor structure of the primary production with final-state
interactions in the \(p\eta\) and \(\bar{p}\eta\) subsystems, using the
chiral unitary approach. The low-energy \(S_{11}\) dynamics, including
the dynamically generated \(N^*(1535)\), is incorporated into the complete 
coupled-channel amplitude, and a phenomenological \(N^*(1895)\)
contribution is included coherently. The \(p\eta\) and
\(\bar{p}\eta\) invariant-mass distributions are well described, and
the \(p\bar{p}\) distribution is predicted as a kinematic reflection
of the \(\eta N\) dynamics. The results support the dynamical
interpretation of the \(N^*(1535)\) and suggest possible additional
\(S_{11}\) strength near \(1.9\)~GeV.
\end{abstract}

\maketitle
\section{Introduction}\label{sec:introduction}

The decay $\eta_c\to p\bar p\eta$ has recently been observed by the
BESIII Collaboration through the radiative process
$\psi(3686)\to\gamma\eta_c$, followed by
$\eta_c\to p\bar p\eta$~\cite{BESIII:2026kdm}. This observation adds a
new baryon--antibaryon--pseudoscalar final state to the still limited
information on baryonic decays of spin-singlet charmonia.  
A substantial part of the $\eta_c$ hadronic width is distributed among exclusive modes. Many of these modes are still poorly measured. Therefore, new decay channels are valuable for clarifying the hadronization mechanisms of the $\eta_c$ decay~\cite{ParticleDataGroup:2026mpi}.
Related three-body baryonic decays,
such as $\eta_c\to p\bar p\pi^0$ and
$\eta_c\to K^+\Lambda\bar p+\mathrm{c.c.}$, have previously been
reported~\cite{BESIII:2012urf,Belle:2018ies}. The presence of an $\eta$
meson in the newly observed channel is particularly interesting because
it gives direct access to the $S$-wave $\eta N$ interaction.

The three-body final state provides the three two-body invariant-mass
distributions
$M_{\eta p}$, $M_{\eta\bar p}$, and $M_{p\bar p}$. The first two probe
the same meson--nucleon dynamics in charge-conjugate subsystems, whereas
the $p\bar p$ distribution can contain kinematic reflections of the same
subsystem dynamics. The structures reported in the $p\eta$ and
$\eta\bar p$ spectra are consistent with an important contribution from
the $N^*(1535)$ and the corresponding antiparticle~\cite{BESIII:2026kdm}. This
resonance couples strongly to the $\eta N$ channel and plays a central
role in the near-threshold $S_{11}$ amplitude. In the chiral unitary
approach, it is generated dynamically by the coupled-channel interaction
of the $\pi N$, $\eta N$, $K\Lambda$, and $K\Sigma$ states
~\cite{Kaiser:1996js,Nacher:1999vg,Inoue:2001ip,Nieves:2001wt,Kolomeitsev:2003kt}.

A dynamically generated resonance should not, in general, be separated from the associated coupled-channel background and represented by an isolated Breit--Wigner term. The line shape of this state in a production process depends on the primary weights of the different meson--baryon channels, the channel-dependent loop functions, and the coherent interference between the direct and rescattering mechanisms.
The same $S_{11}$ partial wave
also contains higher-energy strength commonly associated with the
$N^*(1650)$~\cite{Bruns:2019fwi}. Consequently, the structures observed
in $\eta_c\to p\bar p\eta$ should be analyzed in terms of the complete 
coupled-channel amplitude rather than as an incoherent sum of individual
low-lying resonances.

The possible role of the higher-mass $N^*(1895)$, with
$J^P=1/2^-$, is also of interest. Coupled-channel studies involving
pseudoscalar- and vector-meson channels have associated this state with
dynamically generated poles and have proposed a two-pole structure
~\cite{Khemchandani:2013nma,Khemchandani:2020exc}. The relevance of this resonance to
$\eta N$ and $\eta' N$ production has also been discussed in
phenomenological and partial-wave analyses
~\cite{Khemchandani:2020exc,Anisovich:2017pox}. Since the
strangeness-zero chiral amplitude employed below is designed primarily
to describe the lower-energy $S_{11}$ region, any additional strength
near $1.9$~GeV is represented in the present calculation by
phenomenological $N^*(1895)$ contributions in the two charge-conjugate
subsystems. In accordance with the numerical implementation, these terms
are added coherently at the amplitude level and can therefore interfere
with both the direct source and the coupled-channel rescattering
contributions.

Three-body charmonium decays have proved useful for studying such
final-state-interaction effects and the dynamical origin of baryon
resonances. Applications to
$J/\psi\to\Lambda\bar\Lambda\pi^0$,
$J/\psi\to\Lambda\pi\bar\Sigma$,
$J/\psi\to\Lambda\bar\Sigma^0\eta$,
$J/\psi\to\Xi^0\bar\Lambda K^0$, $J/\psi\to\bar\Lambda\pi\Sigma$, and
$\psi(3686)\to\Lambda\bar\Lambda\eta$ have illustrated how the SU(3)
flavor structure of the primary source and the subsequent
meson--baryon rescattering jointly determine the measured line shapes
~\cite{Ren:2026ila,Lyu:2026ums,Dai:2026zqn,Lyu:2026rsm,He:2026mkf,Ikeno:2026zpv}.
These studies are based on the chiral unitary approach, in which the
interaction of the pseudoscalar-meson matrix $P$ with the baryon octet is
derived from the chiral Lagrangian and then unitarized through the
coupled-channel scattering equation
~\cite{Oset:1997it,Oset:2001cn,Garcia-Recio:2002yxy,
Kolomeitsev:2003kt,Oller:2006jw}. This framework has been successfully used to describe several low-lying baryon resonances as dynamically
generated states. It also shows that the line shape observed in a
production process can be strongly affected by the interference
between the tree-level contribution and the meson--baryon
rescattering terms.
The latter reaction is particularly close in organization to the present
one: the leading single-trace SU(3) structures are first constructed,
and the primary meson--baryon channels are then dressed by
coupled-channel final-state interactions. The spin structure must,
however, be treated differently because the $\eta_c$ has spin zero and
carries no polarization four-vector.

Previous studies of related charmonium decays have also demonstrated
the importance of intermediate nucleon resonances. In the
$J/\psi\to p\bar p\eta$ reaction, the $N^*(1535)$ and
$N^*(1650)$ were found to give important contributions to the
$\eta N$ invariant-mass distribution~\cite{BES:2001gvq}. The
$\eta_c\to p\bar p\eta$ decay provides a complementary environment
for studying similar dynamics in a spin-singlet charmonium decay. An
important difference is that the $\eta_c$ has spin zero and does not
carry a polarization four-vector. The spin structure of the primary
production vertex must therefore be treated differently from that of
a vector-charmonium decay.

In this work, we construct the primary production weights for
$\eta_c\to B\bar B P$ in the strangeness-zero sector and evaluate the
final-state interactions in the $p\eta$ and $\eta\bar p$ subsystems.
The low-energy $S_{11}$ dynamics, including the structure associated
with the $N^*(1535)$, is incorporated through the complete chiral-unitary
$\pi N$--$\eta N$--$K\Lambda$--$K\Sigma$ amplitude, rather than through
separate Breit--Wigner parametrizations. We then add the phenomenological
$N^*(1895)$ contribution described above and calculate the
$p\eta$, $\eta\bar p$, and $p\bar p$ invariant-mass distributions from
a common coherent three-body amplitude. The $p\bar p$ spectrum is a
projection of this amplitude; no independent $p\bar p$ rescattering
mechanism is included in the present model,
since meson baryon amplitudes, including resonances, have much larger strength than baryon baryon ones.


\section*{Formalism}\label{sec2}

\subsection{Flavor structure for \(\eta_c \to p\bar{p}\eta\)}

We study the decay \(\eta_c \to p\bar{p}\eta\). The \(\eta_c\) is a
\(c\bar c\) state and hence a singlet under light-flavor SU(3) in the
\(u,d,s\) sector. The final state consists of a baryon \(p\), an
antibaryon \(\bar p\), and a pseudoscalar meson \(\eta\). The reaction
proceeds through the production of a meson--baryon pair and the
subsequent final-state interaction, which leads to the \(p\eta\) and
\(\bar p\eta\) invariant-mass distributions. We follow the formalism
developed in Refs.
~\cite{He:2026mkf,Dai:2026zqn,Lyu:2026rsm,Ren:2026ila,Lyu:2026ums,Ikeno:2026zpv}

We label the final particles as \(p(1)\), \(\eta(2)\), and
\(\bar p(3)\), and define
$
 M_{12}=M_{p\eta}, 
 M_{23}=M_{\eta\bar p},
 M_{13}=M_{p\bar p}.
$
To determine the flavor structure, we introduce the pseudoscalar-meson
matrix \(P\), the baryon octet $B$, and the antibaryon octet
\(\bar B\). These matrices are written as
\begin{equation}
\resizebox{0.38\linewidth}{!}{$
P=
\begin{pmatrix}
\dfrac{1}{\sqrt{2}}\pi^0 + \dfrac{1}{\sqrt{3}}\eta & \pi^+ & K^+ \\
\pi^- & -\dfrac{1}{\sqrt{2}}\pi^0 + \dfrac{1}{\sqrt{3}}\eta & K^0 \\
K^- & \bar K^0 & -\dfrac{1}{\sqrt{3}}\eta
\end{pmatrix}
$}
\label{eq:Pmatrix}
\end{equation}
\begin{equation}
\resizebox{0.38\linewidth}{!}{$
B=
\begin{pmatrix}
\dfrac{1}{\sqrt{2}}\Sigma^0 + \dfrac{1}{\sqrt{6}}\Lambda & \Sigma^+ & p \\
\Sigma^- & -\dfrac{1}{\sqrt{2}}\Sigma^0 + \dfrac{1}{\sqrt{6}}\Lambda & n \\
\Xi^- & \Xi^0 & -\dfrac{2}{\sqrt{6}}\Lambda
\end{pmatrix}
$}
\label{eq:Bmatrix}
\end{equation}
\begin{equation}
\resizebox{0.38\linewidth}{!}{$
\bar B=
\begin{pmatrix}
\dfrac{1}{\sqrt{2}}\bar\Sigma^0 + \dfrac{1}{\sqrt{6}}\bar\Lambda & \bar\Sigma^+ & \bar\Xi^+ \\
\bar\Sigma^- & -\dfrac{1}{\sqrt{2}}\bar\Sigma^0 + \dfrac{1}{\sqrt{6}}\bar\Lambda & \bar\Xi^0 \\
\bar p & \bar n & -\dfrac{2}{\sqrt{6}}\bar\Lambda
\end{pmatrix}
$}
\label{eq:Bbarmatrix}
\end{equation}

Here, in \(P\), we have considered the \(\eta\)--\(\eta'\) mixing of
Ref.~\cite{Bramon:1992kr}. We can write the SU(3) scalar structures in
terms of the traces \(\langle\bar B B P\rangle\),
\(\langle\bar B P B\rangle\),
\(\langle\bar B B\rangle\langle P\rangle\),
\(\langle\bar B P\rangle\langle B\rangle\), and
\(\langle B P\rangle\langle\bar B\rangle\). Since $\widetilde B$ and
\(\bar B\) have zero trace, only
\(\langle\bar B B P\rangle\), \(\langle\bar B P B\rangle\), and
\(\langle\bar B B\rangle\langle P\rangle\) survive. 
However, the \(\langle\bar B B\rangle\langle P\rangle\) structure is suppressed by large-\(N_c\)  counting~\cite{Manohar:1998xv,Abreu:2023yvf,Dai:2026zqn}, and is further reduced because if no  \(\eta\)--\(\eta'\) mixing is used, $P$
has also zero trace and only the single trace terms remain.
Thus, we retain the
two single-trace structures \(\langle\bar B P B\rangle\) and
\(\langle\bar B B P\rangle\), which are equivalent to those used in
Refs.~\cite{He:2026mkf,Dai:2026zqn,Lyu:2026rsm,Ren:2026ila,Lyu:2026ums,Ikeno:2026zpv}.

We assign the weights \(\widetilde A\) and \(\widetilde B\) to these
two structures. At this stage, the two coefficients specify the flavor
part of the primary source,
\begin{equation}
 \mathcal F_{\eta_c\bar B B P}
 =\widetilde A\,\langle\bar B P B\rangle
 +\widetilde B\,\langle\bar B B P\rangle.
 \label{eq:flavor-source}
\end{equation}
The spin structure is introduced separately below. The two coefficients
can also be written in terms of the symmetric and antisymmetric SU(3)
couplings,
\begin{equation}
 \widetilde A=\widetilde D+\widetilde F,
 \qquad
 \widetilde B=\widetilde D-\widetilde F,
 \label{eq:DFrelation}
\end{equation}
where \(\widetilde D\) and \(\widetilde F\) denote the standard
symmetric and antisymmetric couplings~\cite{deSwart:1963pdg,He:2026mkf}.

We shall take into account the meson--baryon and
antimeson--antibaryon interactions. For the topology in which the
antiproton is the spectator, the two flavor traces give
\begin{align}
\left.\langle\bar B P B\rangle\right|_{\bar p\,\mathrm{spec.}}
={}&\bar p\left[
 \left(\frac{\pi^0}{\sqrt{2}}+\frac{\eta}{\sqrt{3}}\right)p
 +\pi^+n
 -\sqrt{\frac{2}{3}}K^+\Lambda
 \right],
\label{eq:BPB-pbarspec}
\\
\left.\langle\bar B B P\rangle\right|_{\bar p\,\mathrm{spec.}}
={}&\bar p\left[
 -\frac{\eta p}{\sqrt{3}}
 +\frac{K^+\Lambda}{\sqrt{6}}
 +\frac{K^+\Sigma^0}{\sqrt{2}}
 +K^0\Sigma^+
 \right].
\label{eq:BBP-pbarspec}
\end{align}
The charge-conjugate terms, with the proton as spectator, are
\begin{align}
\left.\langle\bar B P B\rangle\right|_{p\,\mathrm{spec.}}
={}&p\left[
 \left(\frac{\pi^0}{\sqrt{2}}+\frac{\eta}{\sqrt{3}}\right)\bar p
 +\pi^-\bar n
 -\sqrt{\frac{2}{3}}K^-\bar\Lambda
 \right],
\label{eq:BPB-pspec}
\\
\left.\langle\bar B B P\rangle\right|_{p\,\mathrm{spec.}}
={}&p\left[
 -\frac{\eta\bar p}{\sqrt{3}}
 +\frac{K^-\bar\Lambda}{\sqrt{6}}
 +\frac{K^-\bar\Sigma^0}{\sqrt{2}}
 +\bar K^0\bar\Sigma^-
 \right].
\label{eq:BBP-pspec}
\end{align}
Consequently, in the phase convention of
Eqs.~\eqref{eq:Pmatrix}--\eqref{eq:Bbarmatrix}, the primary weights for
the meson--baryon channels are
\begin{align}
 h_{\pi^0p}&=\frac{\widetilde A}{\sqrt{2}},
 &h_{\pi^+n}&=\widetilde A,
 &h_{\eta p}&=\frac{\widetilde A-\widetilde B}{\sqrt{3}},
 \notag\\
 h_{K^+\Lambda}&=\frac{-2\widetilde A+\widetilde B}{\sqrt{6}},
 &h_{K^+\Sigma^0}&=\frac{\widetilde B}{\sqrt{2}},
 &h_{K^0\Sigma^+}&=\widetilde B,
\label{eq:primary-weights}
\end{align}
and the charge-conjugate channels carry the corresponding weights.
The relevant strangeness-zero coupled channels are therefore
\(\pi N\), \(\eta N\), \(K\Lambda\), and \(K\Sigma\).

\subsection{Rescattering mechanisms}

\begin{figure}[H]
\centering
\IfFileExists{eta_c_formalism_diagrams_PLB_style.pdf}{%
\includegraphics[width=0.69\textwidth]{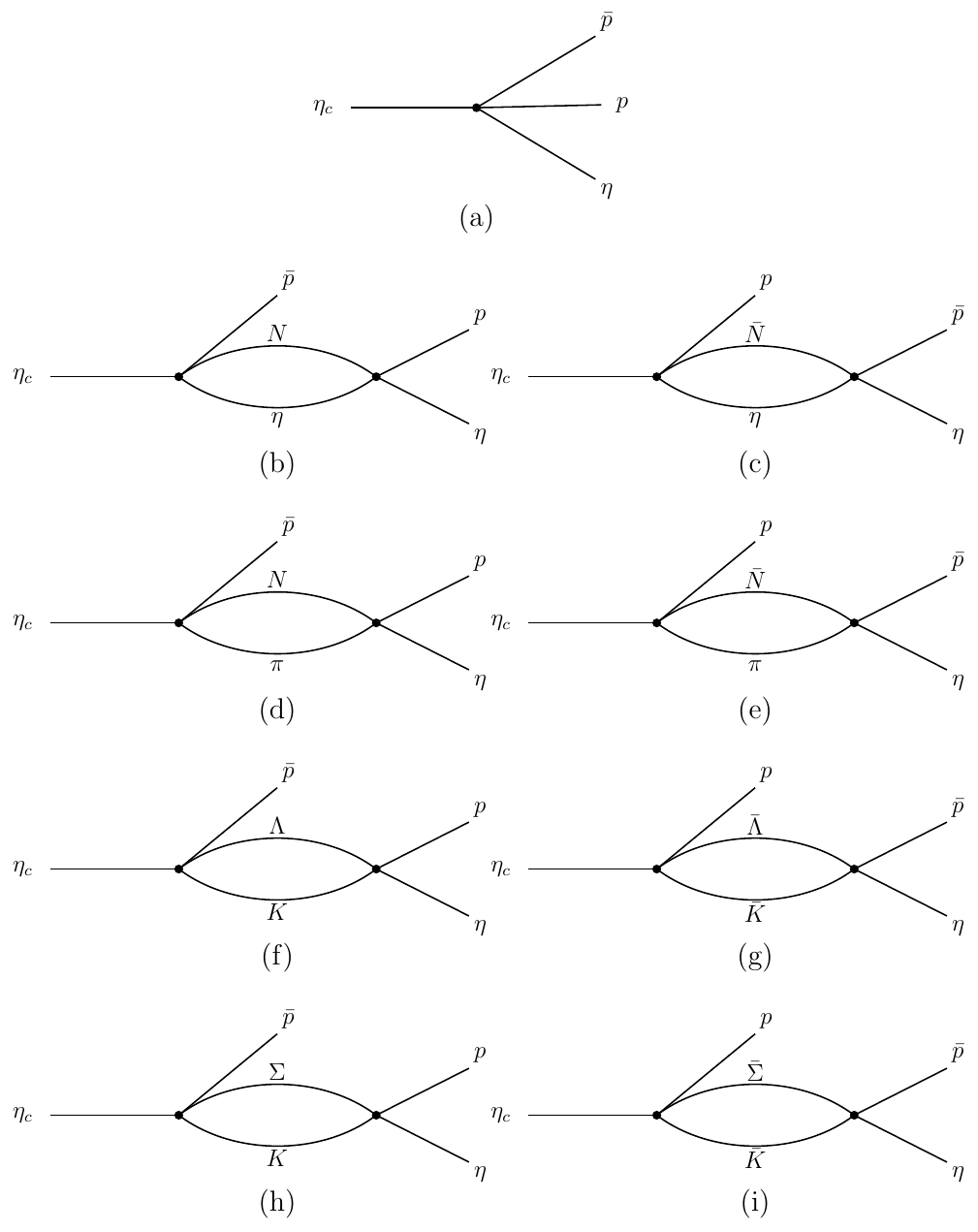}%
}{\fbox{\rule{0pt}{78mm}\rule{0.82\textwidth}{0pt}}}
\caption{Terms contributing to the decay
\(\eta_c\to p\bar p\eta\). Diagram (a) represents the direct tree-level
production mechanism. In diagrams (b), (d), (f), and (h), the
antiproton is a spectator, while the intermediate \(\eta N\),
\(\pi N\), \(K\Lambda\), and \(K\Sigma\) states, respectively,
rescatter into the final \(p\eta\) state. Diagrams (c), (e), (g), and
(i) show the corresponding charge-conjugate mechanisms, in which the
proton is a spectator and an antimeson--antibaryon pair rescatters into
\(\eta\bar p\). The direct tree-level contribution is included
separately from the two rescattering amplitudes defined below.}
\label{fig:diagram}
\end{figure}

The production of \(\eta_c\to p\bar p\eta\) can occur through the
rescattering of intermediate meson--baryon states. The diagrams
contributing to this process are shown in Fig.~\ref{fig:diagram}. The
basic mechanism consists of the \(\eta_c\) decaying into a
baryon--antibaryon--meson state, followed by the final-state interaction
of a meson--baryon or antimeson--antibaryon pair that produces the
\(p\eta\) or \(\bar p\eta\) final state.

Following the formalism of
Refs.~\cite{Ren:2026ila,Lyu:2026ums,Dai:2026zqn,Ikeno:2026zpv,Lyu:2026rsm},
we define two   amplitudes \(t_1\) and \(t_2\) for the rescattering
diagrams in which \(\bar p\) or \(p\) is the spectator, respectively.
The amplitude \(t_1\) corresponds to the case where \(\bar p\) is a
spectator and the meson--baryon pairs interact to produce the \(p\eta\)
final state, while \(t_2\) corresponds to the charge-conjugate case
where \(p\) is a spectator and the antimeson--antibaryon pairs interact
to produce the \(\bar p\eta\) final state.

{

Since the chiral-unitary scattering amplitudes are organized in the
$I=1/2$ sector, we first write the rescattering amplitudes in the
isospin basis.
We next express the reduced isospin amplitudes in terms of the physical
charge channels. For the $\bar p$-spectator topology, the relevant
$I=1/2$, $I_3=1/2$ states are
\begin{align}
|\pi N;\tfrac12,\tfrac12\rangle
&=
-\left(
\sqrt{\tfrac13}\,|\pi^0p\rangle
+
\sqrt{\tfrac23}\,|\pi^+n\rangle
\right),
\notag\\
|\eta N;\tfrac12,\tfrac12\rangle
&=
|\eta p\rangle,
\notag\\
|K\Lambda;\tfrac12,\tfrac12\rangle
&=
|K^+\Lambda\rangle,
\notag\\
|K\Sigma;\tfrac12,\tfrac12\rangle
&=
\sqrt{\tfrac13}\,|K^+\Sigma^0\rangle
+
\sqrt{\tfrac23}\,|K^0\Sigma^+\rangle.
\label{eq:isospin-states-p}
\end{align}
For the resacttering terms
$$  
t_{\pi N \rightarrow \eta N}^{1 / 2},\,
t_{\eta N \rightarrow \eta N}^{1 / 2},\,
t_{K \Lambda \rightarrow \eta N}^{1 / 2},\,
t_{K \Sigma \rightarrow \eta N}^{1 / 2},\,
$$
we need these states and  write the amplitudes  
in terms of charge basis with total charge zero to take advantage of the results of Ref.~\cite{Inoue:2001ip}, as 
{
\begin{align}
& t_{\pi N \rightarrow \eta N}^{1 / 2}= \sqrt{\frac{1}{3}} t_{\pi^0 n \rightarrow \eta n}-\sqrt{\frac{2}{3}} t_{\pi^{-} p \rightarrow \eta n}, \\
& t_{\eta N \rightarrow \eta N}^{1 / 2}=t_{\eta p \rightarrow \eta p}, \\
& t_{K \Lambda \rightarrow \eta N}^{1 / 2}=t_{K^{+} \Lambda \rightarrow \eta p}, \\
& t_{K \Sigma \rightarrow \eta N}^{1 / 2}= \sqrt{\frac{2}{3}} t_{K^{+} \Sigma^- \rightarrow \eta n}-\sqrt{\frac{1}{3}} t_{K^0 \Sigma^{0} \rightarrow \eta n} .
\end{align}

}

The primary weights of the four isospin channels are
\begin{equation}
\begin{aligned}
 h_{\eta N}
 &=
 \frac{\widetilde A-\widetilde B}{\sqrt{3}},
&
 h_{\pi N}
 &=
 -\sqrt{\frac{3}{2}}\,\widetilde A,
\\
 h_{K\Lambda}
 &=
 \frac{-2\widetilde A+\widetilde B}{\sqrt{6}},
&
 h_{K\Sigma}
 &=
 \sqrt{\frac{3}{2}}\,\widetilde B.
\end{aligned}
\label{eq:isospin-production-weights}
\end{equation}
For the $\bar p$-spectator topology, the rescattering amplitude is
therefore
\begin{align}
t_1(M_{12})={}&
\frac{\widetilde A-\widetilde B}{\sqrt3}
G_{\eta N}(M_{12})
t_{\eta N\to\eta N}^{1/2}(M_{12})
\notag\\
&-\sqrt{\frac32}\,\widetilde A\,
G_{\pi N}(M_{12})
t_{\pi N\to\eta N}^{1/2}(M_{12})
\notag\\
&+\frac{-2\widetilde A+\widetilde B}{\sqrt6}
G_{K\Lambda}(M_{12})
t_{K\Lambda\to\eta N}^{1/2}(M_{12})
\notag\\
&+\sqrt{\frac32}\,\widetilde B\,
G_{K\Sigma}(M_{12})
t_{K\Sigma\to\eta N}^{1/2}(M_{12}).
\label{eq:t1-isospin}
\end{align}
In the charge basis $t_2(M_{23})$  has the same functional as $t_1(M_{12})$  with $t_{i\to\eta N}=t_{\bar i\to\eta\bar N}$.
However, taking advantage that
Charge-conjugation invariance gives
\begin{equation}
 G_{\bar i}(M)=G_i(M),\qquad
 t_{\bar i\to\eta\bar p}(M)=t_{i\to\eta p}(M),
 \label{eq:C-invariance}
\end{equation}
we can write the corresponding expression as
\begin{align}
t_2(M_{23})={}&
\frac{\widetilde A-\widetilde B}{\sqrt3}
G_{\eta N}(M_{23})
t_{\eta N\to\eta N}^{1/2}(M_{23})
\notag\\
&-\sqrt{\frac32}\,\widetilde A\,
G_{\pi N}(M_{23})
t_{\pi N\to\eta N}^{1/2}(M_{23})
\notag\\
&+\frac{-2\widetilde A+\widetilde B}{\sqrt6}
G_{ K\Lambda}(M_{23})
t_{ K\Lambda\to\eta N}^{1/2}(M_{23})
\notag\\
&+\sqrt{\frac32}\,\widetilde B\,
G_{ K\Sigma}(M_{23})
t_{ K\Sigma\to\eta N}^{1/2}(M_{23}).
\label{eq:t2-isospin}
\end{align}
Hence \(t_1\) and \(t_2\) share the same functional form and are evaluated at \(M_{12}\) and \(M_{23}\), respectively.


The amplitudes $t_1$ and $t_2$ defined above contain only the
rescattering contributions. The direct tree-level source is
\begin{equation}
t_{\rm tree}
=
h_{\eta p}
=
\frac{\widetilde A-\widetilde B}{\sqrt3},
\label{eq:tree-amplitude}
\end{equation}
}

\subsection{Scattering amplitudes in the chiral unitary approach}

The scattering amplitudes in Eqs.~\eqref{eq:t1-isospin} and
\eqref{eq:t2-isospin} are obtained from the chiral unitary approach.
We use the coupled-channel Bethe--Salpeter equation following the
formalism of Ref.~\cite{Inoue:2001ip}. The \(T\)-matrix is given by
\begin{equation}
 T=[1-VG]^{-1}V,
 \label{eq:BS}
\end{equation}
where \(V\) is the transition potential derived from the lowest-order
chiral Lagrangian and \(G\) is the diagonal meson--baryon loop
function. The explicit evaluation of the potential gives
\begin{equation}
 V_{ij}(\sqrt{s})
 =-\frac{C_{ij}}{4f_if_j}
 \left[k_i^0(\sqrt{s})+k_j^0(\sqrt{s})\right],
 \label{eq:kernel}
\end{equation}
with
\begin{equation}
 k_i^0(\sqrt{s})=\frac{s+m_i^2-M_i^2}{2\sqrt{s}},
 \label{eq:meson-energy}
\end{equation}
where  \(m_i\) and \(M_i\) are the meson and baryon masses in channel
\(i\). We use different meson decay constants for the different
channels,
\begin{equation}
 f_\pi=93~\mathrm{MeV},\qquad
 f_K=1.22f_\pi,\qquad
 f_\eta=1.30f_\pi,
 \label{eq:decay-constants}
\end{equation}
taken from \(\chi\)PT~\cite{Gasser:1984gg}.
We use the leading Weinberg--Tomozawa kernel and the channel
coefficients of Table~\ref{tab:Cij} from  Ref.~\cite{Inoue:2001ip}. 
\begin{table}[H]
\centering
\caption{The coefficients \(C_{ij}\) entering the interaction
potential.}
\label{tab:Cij}
\begin{tabular}{c|cccccc}
\hline\hline
 & $K^+\Sigma^-$ & $K^0\Sigma^0$ & $K^0\Lambda$
 & $\pi^-p$ & $\pi^0n$ & $\eta n$ \\
\hline
$K^+\Sigma^-$ & $1$ & $-\sqrt{2}$ & $0$ & $0$
 & $-1/\sqrt{2}$ & $-\sqrt{3/2}$ \\
$K^0\Sigma^0$ &  & $0$ & $0$ & $-1/\sqrt{2}$
 & $-1/2$ & $\sqrt{3}/2$ \\
$K^0\Lambda$ &  &  & $0$ & $-\sqrt{3/2}$
 & $\sqrt{3}/2$ & $-3/2$ \\
$\pi^-p$ &  &  &  & $1$ & $-\sqrt{2}$ & $0$ \\
$\pi^0n$ &  &  &  &  & $0$ & $0$ \\
$\eta n$ &  &  &  &  &  & $0$ \\
\hline\hline
\end{tabular}
\end{table}
The loop functions are
regularized with a three-momentum cutoff in the present calculation.
It is expressed as
\begin{equation}
\begin{split}
G_i(s)={}&\int_{|\boldsymbol q|<q_{\max}}
\frac{d^3q}{(2\pi)^3}\,2M_i
\frac{\omega_i(q)+E_i(q)}{2\omega_i(q)E_i(q)}
\\
&\times
\frac{1}{s-[\omega_i(q)+E_i(q)]^2+i\varepsilon},
\end{split}
\label{eq:G_i}
\end{equation}
where
\begin{equation}
 \omega_i(q)=\sqrt{m_i^2+\boldsymbol q^{\,2}},\qquad
 E_i(q)=\sqrt{M_i^2+\boldsymbol q^{\,2}}.
\end{equation}

The complete matrix in Eq.~\eqref{eq:BS} contains the low-energy
\(S_{11}\) dynamics.  We determine the cutoff \(q_{\max}\) by adjusting the peak
position of the \(N^*(1535)\) resonance to
\(\sqrt{s}\simeq1535\)~MeV. With this procedure, we obtain
$
 q_{\max}=1000~\mathrm{MeV}.
$

Alternatively, the pole of the \(T\)-matrix can be located on the
second Riemann sheet by replacing \(G_i\) with
\begin{equation}
 G_i^{\mathrm{II}}(\sqrt{s})
 =G_i(\sqrt{s})
 +i\frac{M_i}{2\pi\sqrt{s}}q_{\mathrm{on}},
 \label{eq:GII}
\end{equation}
for \(\mathrm{Re}(\sqrt{s})>M_i+m_i\), where
\begin{equation}
 q_{\mathrm{on}}
 =\frac{\lambda^{1/2}(s,m_i^2,M_i^2)}{2\sqrt{s}},
 \qquad \mathrm{Im}(q_{\mathrm{on}})>0.
\end{equation}
Using this method, the pole associated with the \(N^*(1535)\) is found
at
$
 \sqrt{s}_p=1532.8361-i\,63.36~\mathrm{MeV},
$
which is consistent with the peak position obtained from the modulus
squared of the amplitude.

\subsection{Additional contributions from higher resonances}

In addition to the low-energy coupled-channel amplitude, the possible
contribution of the higher-mass \(N^*(1895)\) resonance may also be
relevant. The \(N^*(1895)\) has spin-parity \(J^P=1/2^-\) and has been
studied in coupled-channel approaches involving pseudoscalar and vector
mesons~\cite{Khemchandani:2020exc,Khemchandani:2013nma}. 

We include the \(N^*(1895)\) and charge-conjugate
\(\bar N^*(1895)\) contributions coherently at the amplitude level.
For the \(p\eta\) subsystem, the contribution is parametrized as
\begin{equation}
 t_{N^*(1895)}(M_{12})
 =\textcolor{black}{\widetilde C} M_R\frac{1}{M_{12}-M_R+i\Gamma_R/2},
 \label{eq:N1895}
\end{equation}
while the corresponding contribution in the \(\bar p\eta\) subsystem
is written as
\begin{equation}
 t'_{N^*(1895)}(M_{23})
 =\textcolor{black}{\widetilde C} M_R\frac{1}{M_{23}-M_R+i\Gamma_R/2}.
 \label{eq:N1895-prime}
\end{equation}
The same values of $\widetilde C$, \(M_R\), and \(\Gamma_R\) are used for the
two charge-conjugate contributions. We take
\begin{equation}
 M_R=1895~\mathrm{MeV},\qquad
 \Gamma_R=120~\mathrm{MeV},
 \label{eq:N1895-parameters}
\end{equation}
while $\widetilde C$ is determined from the fit to the data. Since the two terms
are added at the amplitude level, they interfere with both the direct
source and the coupled-channel rescattering contributions.

\subsection{Spinor structure of the production amplitude}

The spin structure of the present process is simpler than that of a
vector-charmonium decay. In the vector case, the initial-state
polarization vector allows the structure
\(\bar u_B\gamma^\mu\gamma_5v_{\bar B}\epsilon_\mu\). The \(\eta_c\),
however, has spin zero and does not carry a polarization vector.
{
For the leading local \(\eta_c\to p\bar p\eta\) production vertex,
parity conservation leads to the scalar baryon--antibaryon bilinear,
\begin{equation}
 \bar u_B(\boldsymbol p,r)
 v_{\bar B}(\boldsymbol p',r'),
 \label{eq:scalar-bilinear}
\end{equation}
where \(\boldsymbol p\) and \(\boldsymbol p'\) are the three-momenta
of the baryon and antibaryon, respectively.
}

Keeping the baryon spinors to linear order in the three-momenta, we
write
\begin{equation}
 u_B(\boldsymbol p,r)
 =\begin{pmatrix}
 \chi_r\\[1mm]
 \dfrac{\boldsymbol\sigma\cdot\boldsymbol p}{2m}\chi_r
 \end{pmatrix},
 \qquad
 v_{\bar B}(\boldsymbol p',r')
 =\begin{pmatrix}
 \dfrac{\boldsymbol\sigma\cdot\boldsymbol p'}{2m'}\chi'_{r'}\\[1mm]
 \chi'_{r'}
 \end{pmatrix}.
 \label{eq:nonrelativistic-spinors}
\end{equation}
{
In the Dirac representation,
\begin{equation}
 \gamma^0=\begin{pmatrix}I&0\\0&-I\end{pmatrix}.
 \label{eq:gamma-matrices}
\end{equation}
It follows that
\begin{equation}
 \bar u_B(\boldsymbol p,r)v_{\bar B}(\boldsymbol p',r')
 =\chi_r^\dagger
 \left(
 \frac{\boldsymbol\sigma\cdot\boldsymbol p'}{2m'}
 -\frac{\boldsymbol\sigma\cdot\boldsymbol p}{2m}
 \right)\chi'_{r'}.
 \label{eq:scalar-reduction-leading}
\end{equation}
The production amplitude is therefore linear in the baryon and
antibaryon momenta at the order considered here. For the direct
\(\eta_c\to p\bar p\eta\) source, we have
\begin{equation}
 t_{\rm tree}^{rr'}=
 \frac{\widetilde A-\widetilde B}{\sqrt3}
 \chi_r^\dagger
 \left(
 \frac{\boldsymbol\sigma\cdot\boldsymbol p'}{2m_p}
 -\frac{\boldsymbol\sigma\cdot\boldsymbol p}{2m_p}
 \right)\chi'_{r'}.
 \label{eq:channel-production-amplitude}
\end{equation}
}

{
The two SU(3) flavor structures have the same spin dependence.
In the \(S\)-wave approximation for the intermediate meson--baryon
subsystem, the contributions linear in the internal loop momentum
vanish under angular integration,~\footnote{Eq.~(\ref{eq:loop-angular-average}) is strictly valid in the meson baryon rest frame, in other frames is small, but negligible with respect to the remaining term of Eq.(~\ref{eq:channel-production-amplitude})}
\begin{equation}
 \int\frac{d^3q}{(2\pi)^3}\,\boldsymbol q\,F(\boldsymbol q^{\,2})=0.
 \label{eq:loop-angular-average}
\end{equation}
 The \(S\)-wave
rescattering amplitudes can then be expressed in terms of the reduced
amplitudes \(t_1\) and \(t_2\) in
Eqs.~\eqref{eq:t1-isospin} and \eqref{eq:t2-isospin} as
\begin{align}
 t_1^{rr'}(M_{12})={}&
 \chi_r^\dagger
 \frac{\boldsymbol\sigma\cdot\boldsymbol p'}{2m_p}
 \chi'_{r'}\,t_1(M_{12}),\notag\\
 t_2^{rr'}(M_{23})={}&
 -\chi_r^\dagger
 \frac{\boldsymbol\sigma\cdot\boldsymbol p}{2m_p}
 \chi'_{r'}\,t_2(M_{23}).
 \label{eq:momentum-rescattering}
\end{align}
The phenomenological \(N^*(1895)\) and \(\bar N^*(1895)\)
contributions carry the corresponding momentum structures in the
\(p\eta\) and \(\bar p\eta\) subsystems. We define
\begin{equation}
 \begin{aligned}
 D'(M_{12})={}&\frac{\widetilde A-\widetilde B}{\sqrt3}
 +t_1(M_{12})+t_{N^*(1895)}(M_{12}),\\
 D(M_{23})={}&\frac{\widetilde A-\widetilde B}{\sqrt3}
 +t_2(M_{23})+t'_{N^*(1895)}(M_{23}).
 \end{aligned}
 \label{eq:scalar-total-amplitude}
\end{equation}
The complete decay amplitude is written as
\begin{equation}
 \mathcal M_{rr'}(M_{12},M_{23})
 =\chi_r^\dagger\left[
 D'(M_{12})\frac{\boldsymbol\sigma\cdot\boldsymbol p'}{2m_p}
 -D(M_{23})\frac{\boldsymbol\sigma\cdot\boldsymbol p}{2m_p}
 \right]\chi'_{r'}.
 \label{eq:complete-scalar-amplitude}
\end{equation}
Since the \(\eta_c\) has spin zero, no average over initial
polarizations is required. Summing over the final baryon and
antibaryon spins gives
\begin{equation}
 \begin{split}
 \sum_{r,r'}|\mathcal M_{rr'}|^2={}&
 \frac{2}{4m_p^2}\Big[|D'(M_{12})|^2\boldsymbol p'^{\,2}
 +|D(M_{23})|^2\boldsymbol p^{\,2}\\
 &-2\operatorname{Re}\!\left(D'(M_{12})D^*(M_{23})\right)
 \boldsymbol p\cdot\boldsymbol p'\Big].
 \end{split}
 \label{eq:spin-summed-scalar-amplitude}
\end{equation}
The factor of two follows from the trace over the final Pauli spin
indices. The last term accounts for the coherent interference between
the two momentum-dependent contributions.
}

\subsection{Mass distributions}

Using the Mandl--Shaw normalization for the baryon fields
~\cite{MandlShaw2010}, the double-differential decay width for
\(\eta_c\to p\bar p\eta\) is
\begin{equation}
\begin{split}
 \frac{d^2\Gamma}{dM_{12}\,dM_{23}}
 ={}&\frac{1}{(2\pi)^3}
 \frac{(2m_p)(2m_p)}{32m_{\eta_c}^3}
 \left[\textcolor{black}{\sum_{r,r'}|\mathcal M_{rr'}(M_{12},M_{23})|^2}\right]
 \\
 &\times(2M_{12})(2M_{23}).
\end{split}
\label{eq:double-differential-width}
\end{equation}
The three invariant masses satisfy
\begin{equation}
 M_{12}^2+M_{13}^2+M_{23}^2
 =m_{\eta_c}^2+2m_p^2+m_\eta^2.
 \label{eq:mass-relation}
\end{equation}
{
The three-momenta in Eq.~\eqref{eq:spin-summed-scalar-amplitude}
are evaluated in the \(\eta_c\) rest frame. Their magnitudes are
\begin{equation}
 \begin{aligned}
 |\boldsymbol p|={}&
 \frac{\lambda^{1/2}(m_{\eta_c}^2,m_p^2,M_{23}^2)}{2m_{\eta_c}},\\
 |\boldsymbol p'|={}&
 \frac{\lambda^{1/2}(m_{\eta_c}^2,m_p^2,M_{12}^2)}{2m_{\eta_c}},
 \end{aligned}
 \label{eq:proton-antiproton-momenta}
\end{equation}
where \(\lambda(x,y,z)=x^2+y^2+z^2-2xy-2xz-2yz\).
The corresponding energies are
\begin{equation}
 \begin{aligned}
 E_p&=\frac{m_{\eta_c}^2+m_p^2-M_{23}^2}{2m_{\eta_c}},\\
 E_{\bar p}&=\frac{m_{\eta_c}^2+m_p^2-M_{12}^2}{2m_{\eta_c}}.
 \end{aligned}
 \label{eq:proton-antiproton-energies}
\end{equation}
The scalar product of the two three-momenta can be expressed as
\begin{equation}
 \boldsymbol p\cdot\boldsymbol p'
 =E_p E_{\bar p}-\frac{M_{13}^2-2m_p^2}{2}.
 \label{eq:proton-antiproton-dot}
\end{equation}
}
For fixed \(M_{12}\), the integration limits can be written as
\begin{equation}
 (M_{23}^2)_{\max/\min}
 =(E_2^*+E_3^*)^2-(p_2^*\mp p_3^*)^2,
 \label{eq:Dalitz-limits}
\end{equation}
where
\begin{align}
 E_2^*&=\frac{M_{12}^2-m_p^2+m_\eta^2}{2M_{12}},
&p_2^*&=\sqrt{(E_2^*)^2-m_\eta^2},
\notag\\
 E_3^*&=\frac{m_{\eta_c}^2-M_{12}^2-m_p^2}{2M_{12}},
&p_3^*&=\sqrt{(E_3^*)^2-m_p^2}.
 \label{eq:Dalitz-variables}
\end{align}
The one-dimensional invariant-mass distributions are obtained by
integrating Eq.~\eqref{eq:double-differential-width} over the
corresponding kinematically allowed range. The other distributions
follow by cyclic permutation of the particle labels.

The parameters $\widetilde A$, $\widetilde B$, and $\widetilde C$ are determined by fitting the
BESIII data~\cite{BESIII:2026kdm}. The scattering amplitudes and loop
functions are fixed by the chiral unitary approach described above.



\section{ NUMERICAL Results}\label{sec3}

We perform a simultaneous fit to the BESIII data for the
$p\eta$ and $\bar p\eta$ invariant-mass distributions
~\cite{BESIII:2026kdm}. The three parameters $\widetilde A$,
$\widetilde B$, and $\widetilde C$  are taken as free parameters in the fit. The parameters
$\widetilde A$ and $\widetilde B$ determine the relative strengths
of the two SU(3) flavor structures and, at the same time, fix the
production weights of the intermediate $\pi N$, $\eta N$,
$K\Lambda$, and $K\Sigma$ channels. The parameter $\widetilde C$
determines the strength of the two charge-conjugate $N^*(1895)$
contributions introduced in Eqs.~\eqref{eq:N1895} and
\eqref{eq:N1895-prime}. The meson--baryon scattering amplitudes and
loop functions are fixed by the chiral unitary approach described
in Sec.~\ref{sec2}.

In the normalization convention used in the present calculation,
\textcolor{black}{two sets of fitted parameters are obtained,}
\begin{equation}
\begin{aligned}
\text{Solution A}:\quad
&\widetilde A=6.1019,\quad
 \widetilde B=7.3974,\quad
 \widetilde C=0.58116,\\
\text{Solution B}:\quad
&\widetilde A=29.609,\quad
 \widetilde B=5.8488,\quad
 \widetilde C=0.97916.
\end{aligned}
\label{eq:best-fit-parameters}
\end{equation}
\textcolor{black}{The two fits give}
\begin{equation}
\frac{\chi^2}{\mathrm{d.o.f.}}=
\begin{cases}
1.37, & \text{Solution A},\\
1.39, & \text{Solution B}.
\end{cases}
\label{eq:best-fit-chi2}
\end{equation}
showing that the model provides a good description of the fitted
distributions. \textcolor{black}{The ratios of the two primary-production
coefficients for the two solutions are}
\begin{equation}
\frac{\widetilde A}{\widetilde B}=
\begin{cases}
0.82, & \text{Solution A},\\
5.06, & \text{Solution B}.
\end{cases}
\label{eq:AB-ratio}
\end{equation}
\textcolor{black}{These different ratios indicate that the relative
strengths of the two SU(3) production structures are not uniquely
constrained by the $p\eta$ and $\bar p\eta$ mass distributions alone.
The fitted values of $\widetilde C$ determine the strengths of the
higher-mass contributions and their interference with the other
amplitudes.}

The experimental backgrounds in the $p\eta$ and $\bar p\eta$
channels are added bin by bin to the calculated signal contributions.
They are not included in the complex decay amplitude and therefore
do not interfere with the theoretical signal~\cite{Yang:2026commu}.

\begin{figure}[!htbp]
\centering
\includegraphics[width=0.32\linewidth]{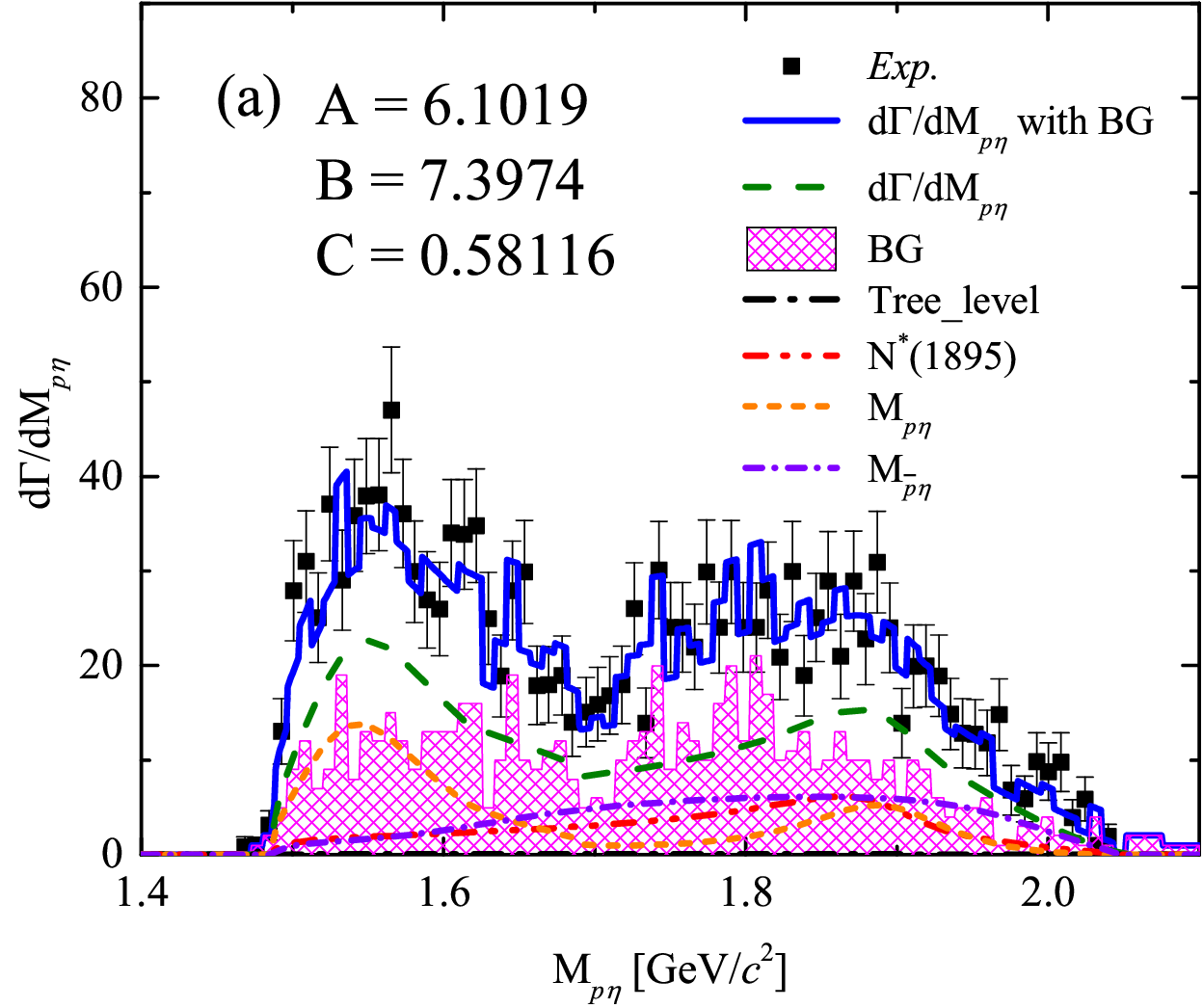}\hfill
\includegraphics[width=0.32\linewidth]{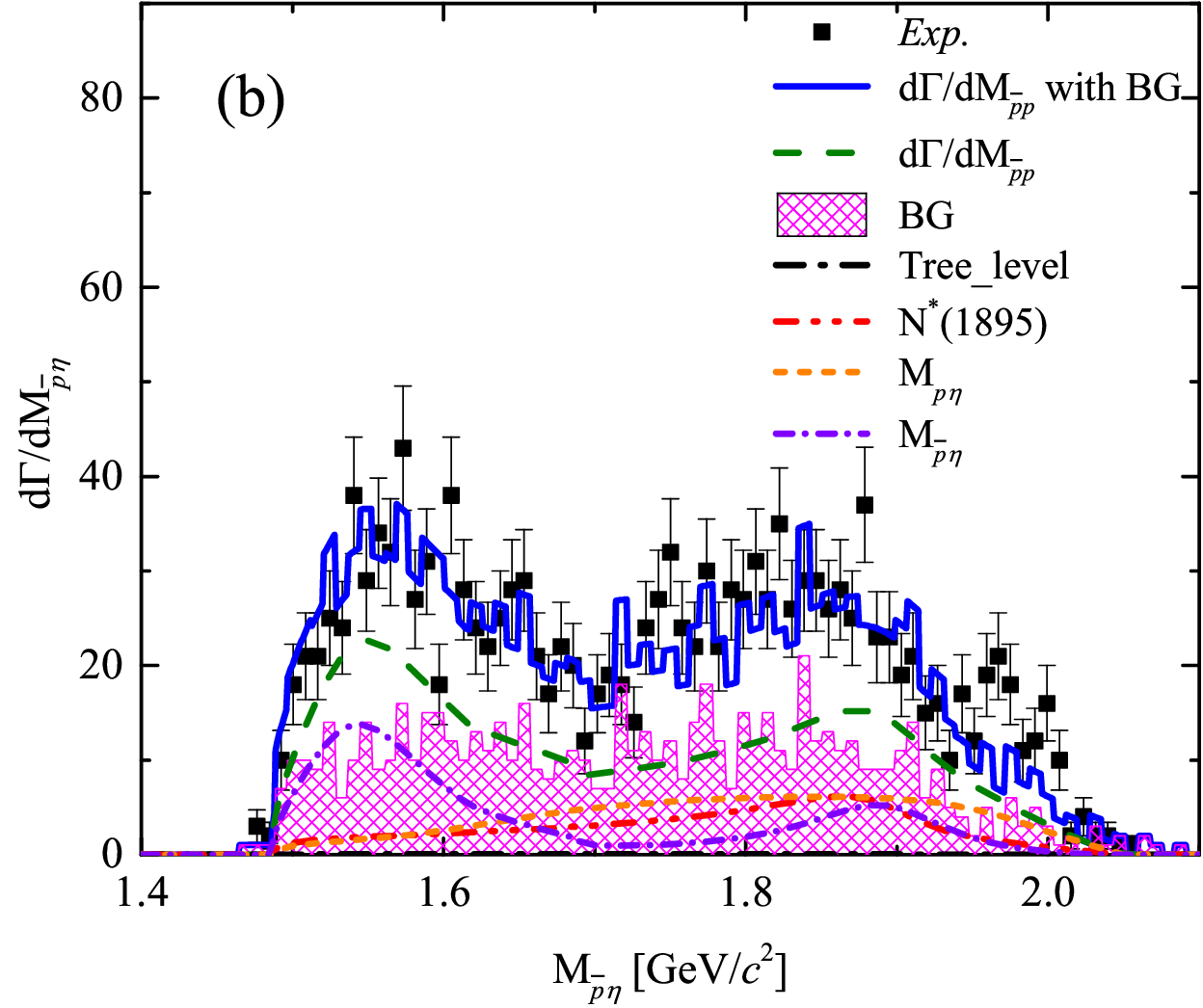}\hfill
\includegraphics[width=0.32\linewidth]{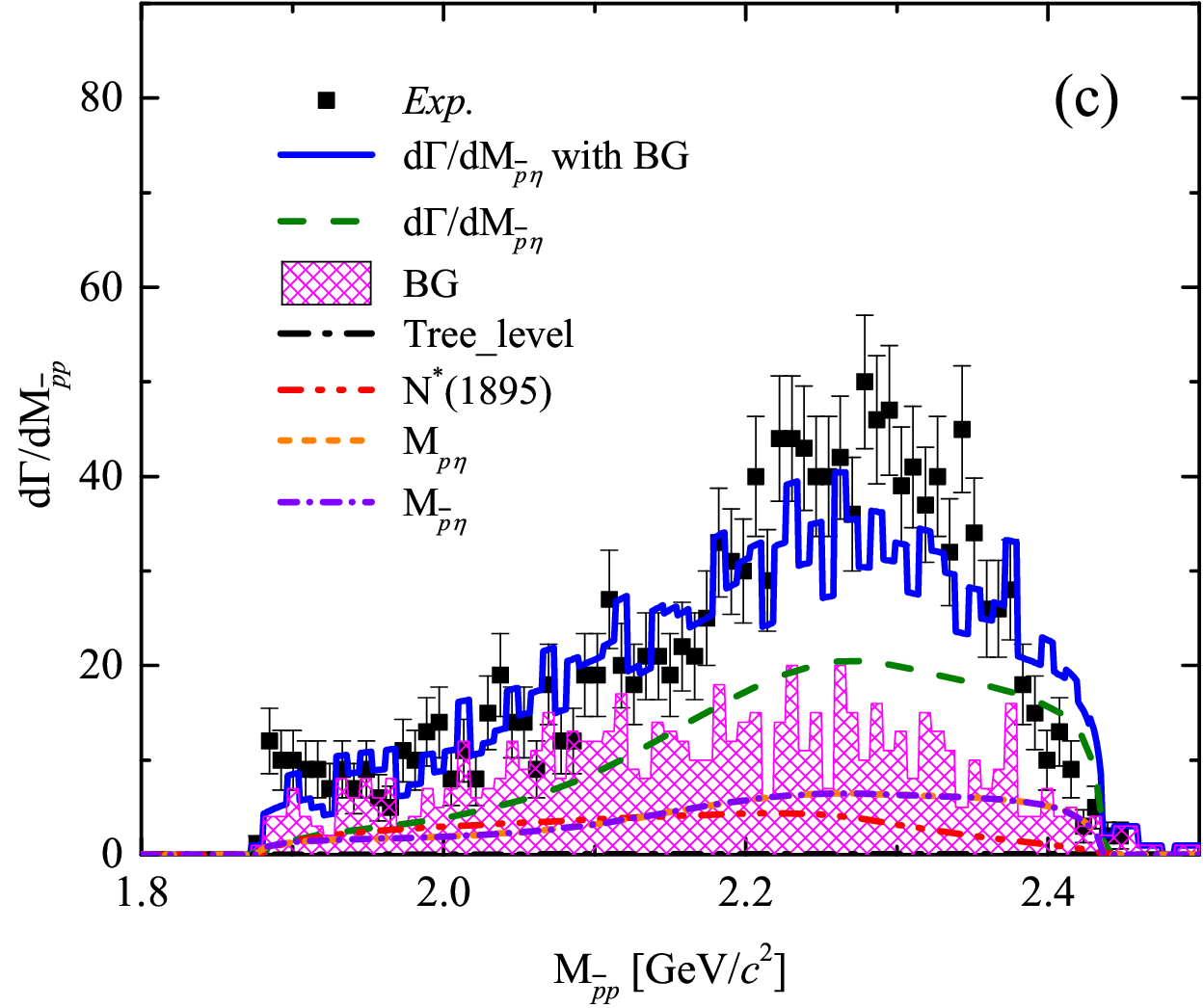}
\caption{Invariant-mass distributions of $p\eta$ (a),
$\bar p\eta$ (b), and $p\bar p$ (c), compared with the BESIII
data~\cite{BESIII:2026kdm}. \textcolor{black}{The fitted parameters
for solution A are $\widetilde A=6.1019$,
$\widetilde B=7.3974$, and $\widetilde C=0.58116$, with
$\chi^2/\mathrm{d.o.f.}=1.37$.} The solid blue curves represent
the total result after adding the experimental background to the
calculated signal contribution. The other curves show the coherent
theoretical signal, the experimentally determined background, the
tree-level contribution, \textcolor{black}{the phenomenological
$N^*(1895)$ and $\bar N^*(1895)$ contributions, and the individual
$M_{p\eta}$ and $M_{\bar p\eta}$ branches, which contain the
rescattering and higher-mass contributions.} The $p\bar p$
distribution is calculated with the parameters obtained from the
simultaneous fit to the $p\eta$ and $\bar p\eta$ distributions.
The individual mechanism curves are obtained by retaining the
corresponding amplitudes separately and therefore do not add
directly to the coherent total result. The fit is performed to
the experimental event counts.}
\label{fig:mass-distributions_A}
\label{fig:mass-distributions}
\end{figure}

\begin{figure}[!htbp]
\centering
\includegraphics[width=0.32\linewidth]{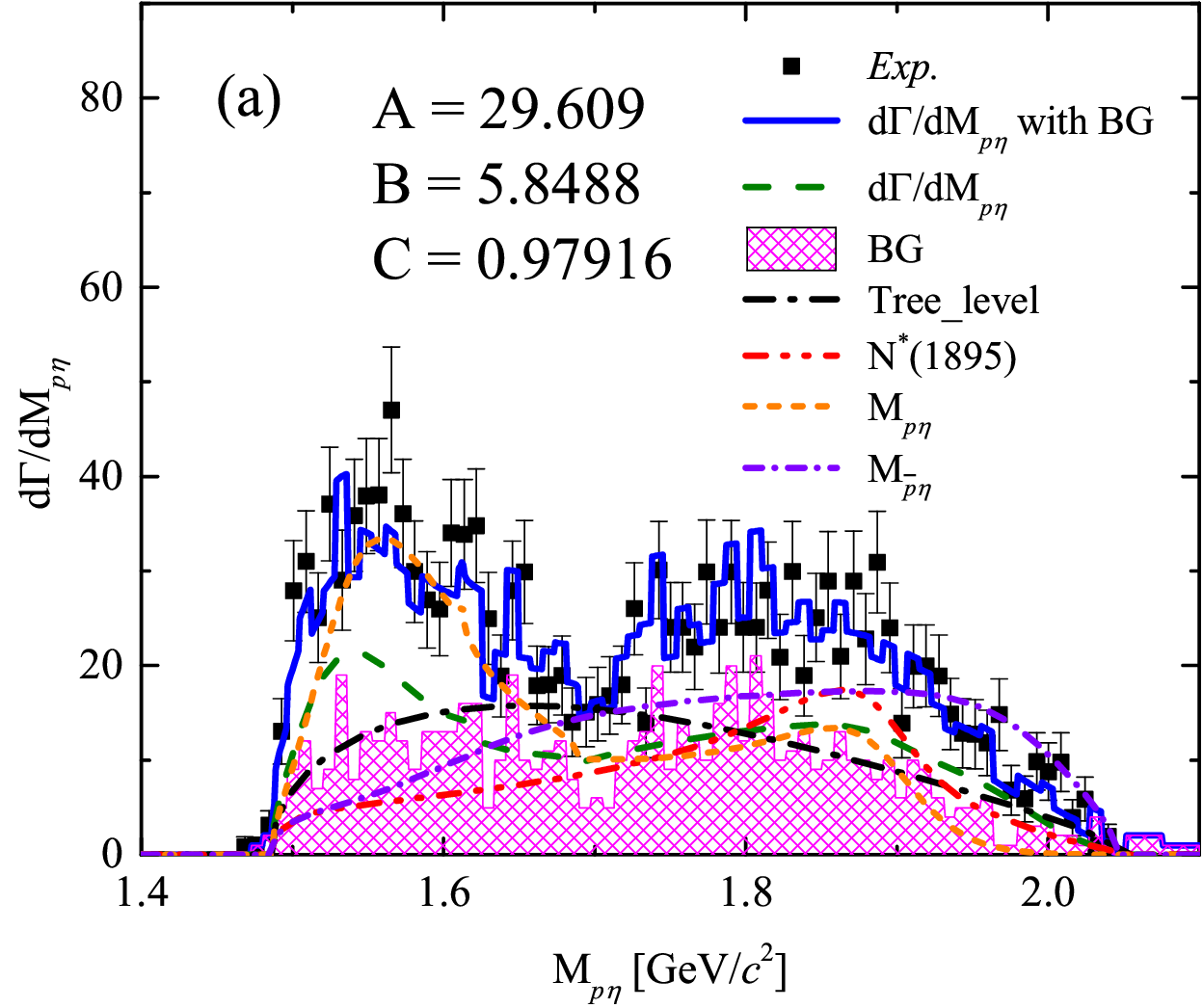}\hfill
\includegraphics[width=0.32\linewidth]{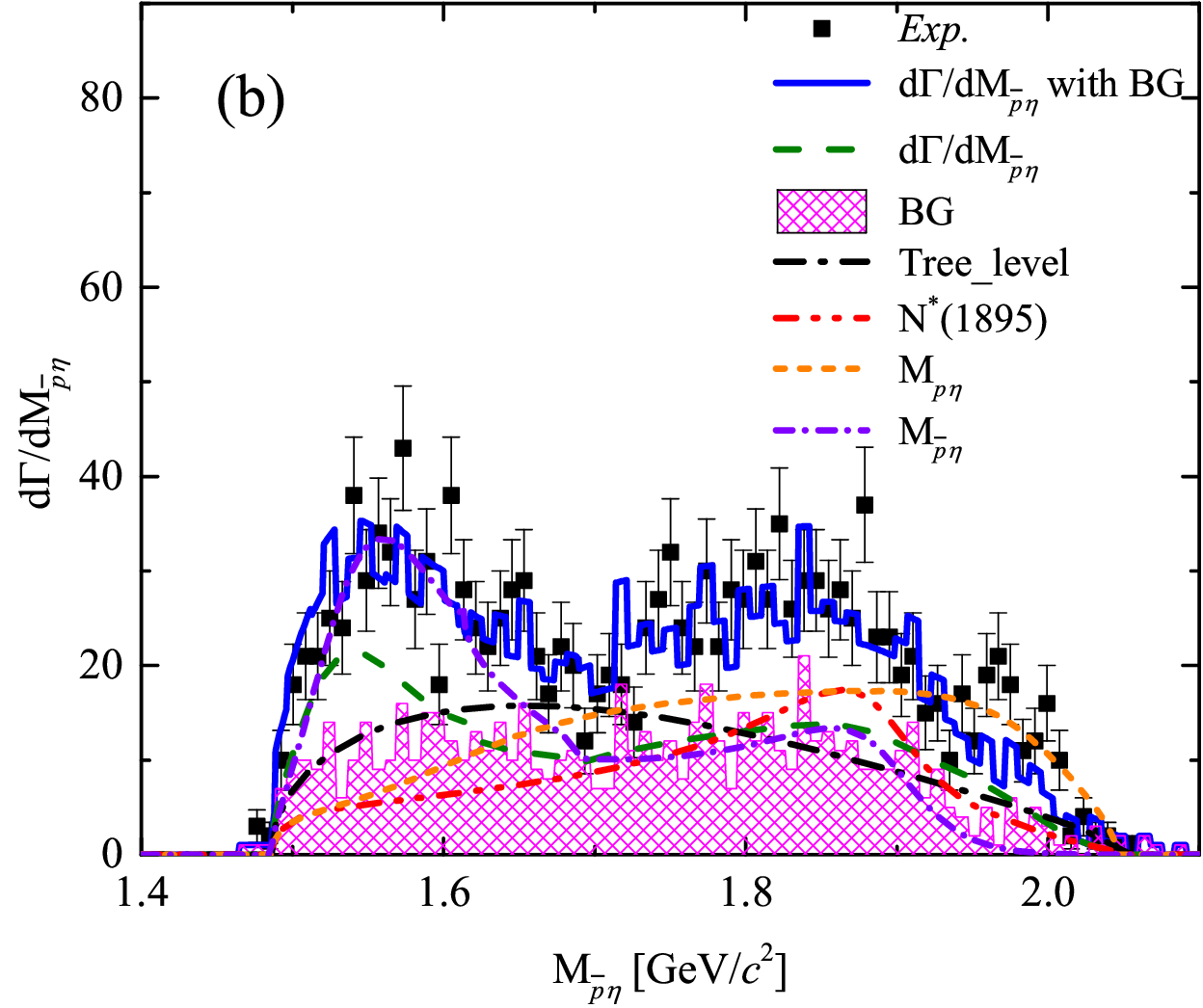}\hfill
\includegraphics[width=0.32\linewidth]{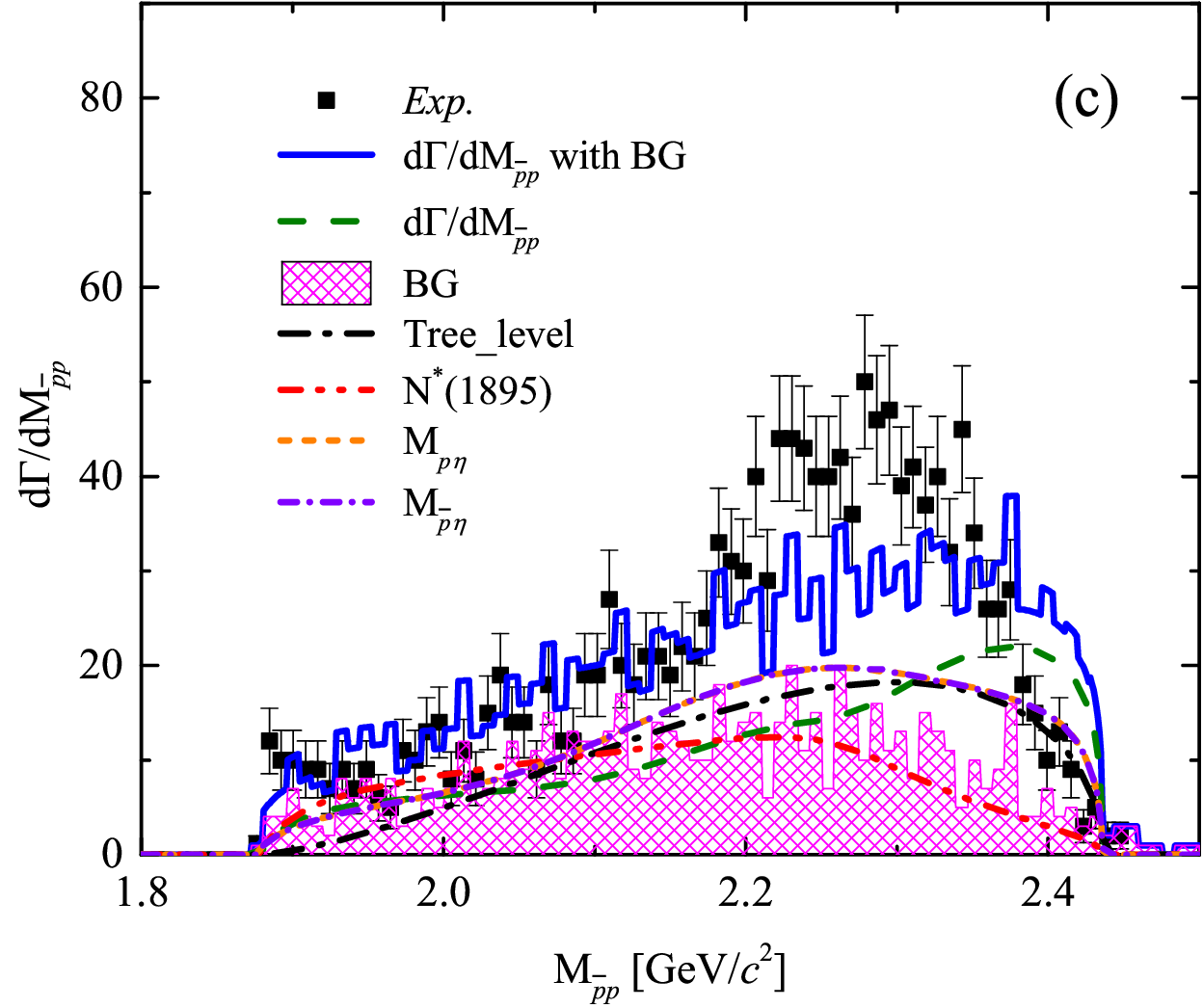}
\caption{\textcolor{black}{Same as Fig.~\ref{fig:mass-distributions_A},
but for solution B, with $\widetilde A=29.609$,
$\widetilde B=5.8488$, and $\widetilde C=0.97916$, giving
$\chi^2/\mathrm{d.o.f.}=1.39$.}}
\label{fig:mass-distributions_B}
\end{figure}

\textcolor{black}{The results obtained with solutions A and B are shown
in Figs.~\ref{fig:mass-distributions_A} and
\ref{fig:mass-distributions_B}, respectively.} The $p\eta$ and
$\bar p\eta$ distributions are fitted simultaneously, while the
$p\bar p$ distribution is calculated with the same parameter set.
The latter distribution therefore provides an additional test of the
reaction mechanism determined from the two $\eta N$ subsystems.
\textcolor{black}{Both solutions give comparable descriptions of the
fitted distributions, with solution A yielding a slightly smaller
$\chi^2$. The $p\bar p$ mass distribution is better described by
solution A.}

To better understand the role of the different mechanisms, we show
the tree-level, \textcolor{black}{higher-mass, and individual $p\eta$
and $\bar p\eta$ contributions} separately in
\textcolor{black}{Figs.~\ref{fig:mass-distributions_A} and
\ref{fig:mass-distributions_B}}. The tree-level term originates
from the direct $\eta_c\to p\bar p\eta$ production vertex and gives
a smooth contribution over the available invariant-mass region. The
tree-level term alone cannot reproduce the structures observed in the
$p\eta$ and $\bar p\eta$ distributions.
\textcolor{black}{In solution A, the direct tree-level contribution is
relatively weak, whereas in solution B it becomes significantly larger.}

The complete chiral-unitary rescattering contribution generates the
main low-energy structure in the two $\eta N$ invariant-mass spectra.
A pronounced enhancement appears in the region of the $N^*(1535)$,
which couples strongly to the $\eta N$ channel. In the present
framework, this enhancement is generated through the coupled-channel
interactions of $\pi N$, $\eta N$, $K\Lambda$, and $K\Sigma$. The
coupled-channel calculation also produces a broader higher-energy
behavior in the $S_{11}$ partial wave.

The phenomenological $N^*(1895)$ and $\bar N^*(1895)$ terms
contribute mainly to the higher $p\eta$ and $\bar p\eta$
invariant-mass regions, respectively. Since the two terms are added
at the amplitude level, they interfere with both the direct-production
term and the coupled-channel rescattering amplitude. The fitted value
of $\widetilde C$ therefore determines both the magnitude of the
higher-mass contributions and \textcolor{black}{their interference with
the other contributions}. \textcolor{black}{Their net effect on the
invariant-mass distributions is determined by these interference
contributions, particularly in the region around $1.9$~GeV.}

\begin{figure}[!htbp]
\centering
\includegraphics[width=0.485\linewidth]{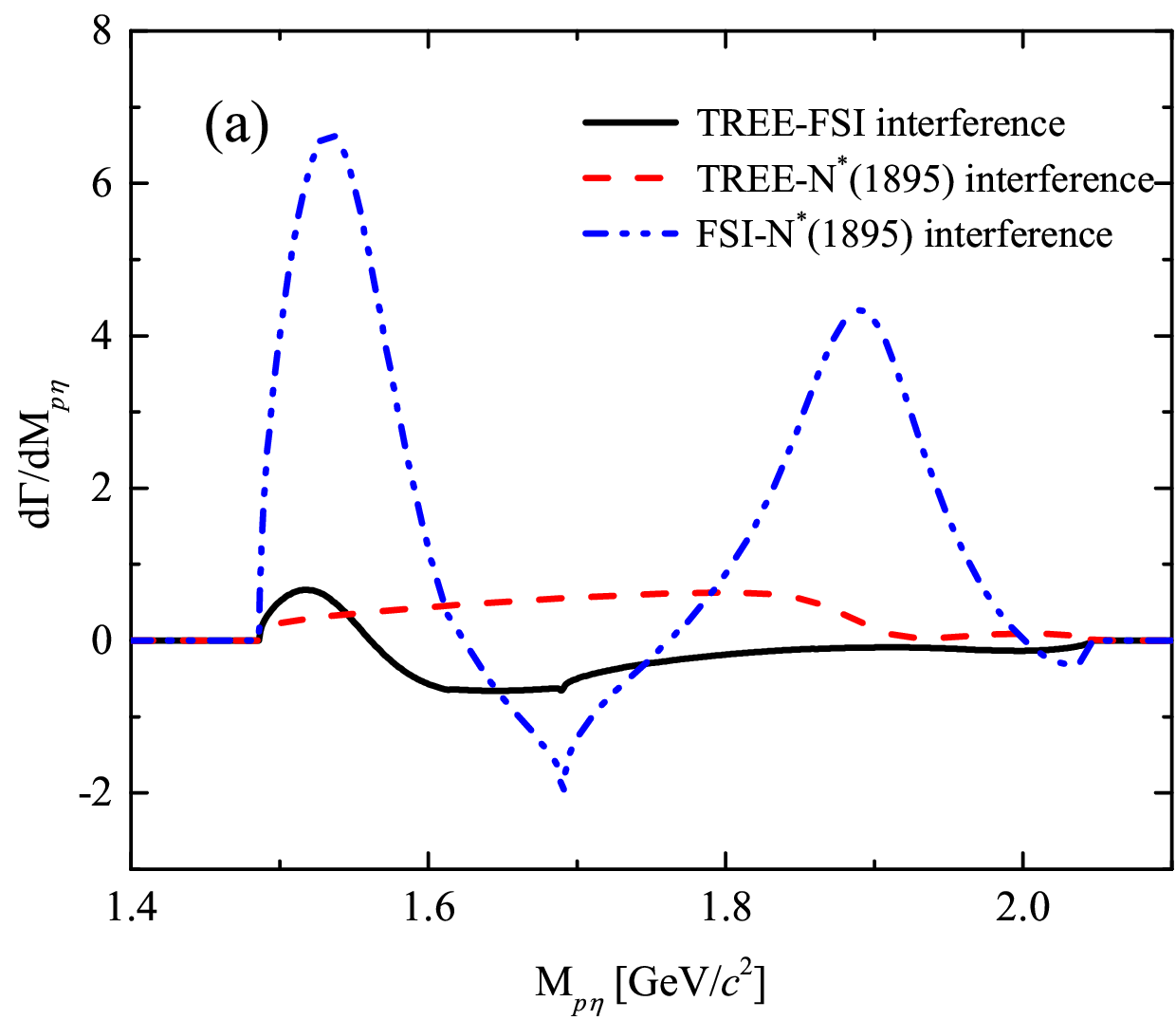}\hfill
\includegraphics[width=0.485\linewidth]{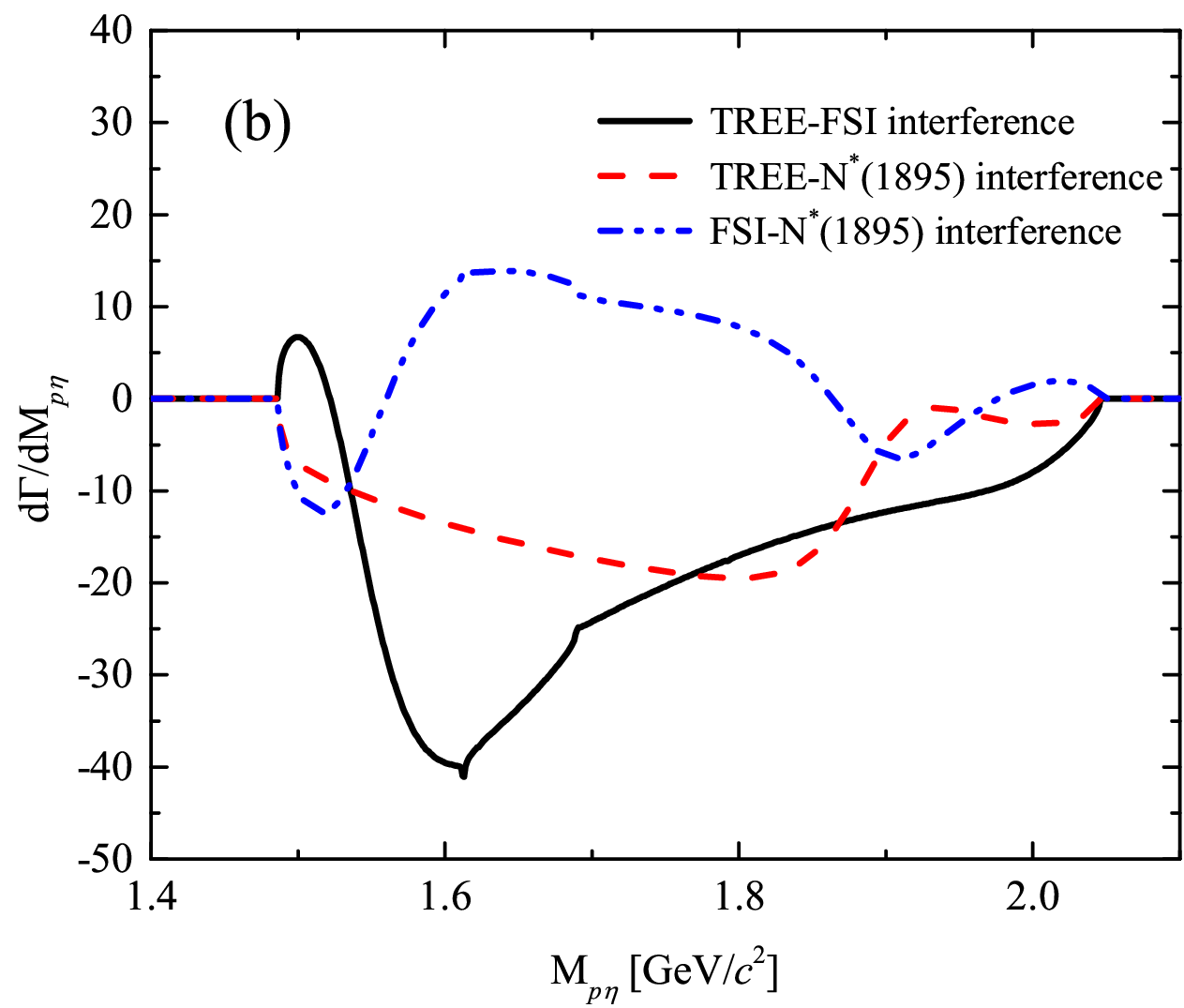}\hfill
\caption{\textcolor{black}{Interference contribution to the $p\eta$
invariant-mass distributions for solution A
[(a)] and solution B [(b)]. The corresponding contributions for $\bar p\eta$ are identical.
The curves represent the interference between the tree-level
and rescattering amplitudes, the tree-level and $N^*(1895)$
amplitudes, and the rescattering and $N^*(1895)$ amplitudes.}}
\label{fig:interference}
\end{figure}

\textcolor{black}{The interference contributions for the two solutions
are shown in Fig.~\ref{fig:interference}.
In solution A, the interference terms involving the tree-level
amplitude remain relatively small, whereas the FSI--$N^*(1895)$
interference gives the most prominent contribution and changes sign
across the invariant-mass region.
In solution B, the TREE--FSI and TREE--$N^*(1895)$ interference
terms are predominantly destructive over a broad mass region.
The FSI--$N^*(1895)$ interference, however, is constructive in the
intermediate-mass region and partially compensates these negative
contributions.
The different mass-dependent interference patterns allow the two
solutions to produce similar total distributions despite their
different production strengths.}

The $p\bar p$ invariant-mass distribution, shown in
\textcolor{black}{Figs.~\ref{fig:mass-distributions_A}(c) and
\ref{fig:mass-distributions_B}(c)}, exhibits a broad structure
that arises from the kinematic reflection of the structures produced
in the $p\eta$ and $\bar p\eta$ channels. The $p\bar p$
distribution provides a prediction or cross-check of the fitted
three-body amplitude. No independent $p\bar p$ final-state interaction
is introduced in the present calculation.
\textcolor{black}{The first solution provides a better description of
the $p\bar p$ distribution, favoring this parameter set when all
three mass distributions are considered.}

The experimental background also plays an important role in the
comparison with the data. In particular, the background contribution
accounts for part of the {event counts} in regions
where the theoretical signal becomes small. After the experimental
background is added to the calculated signal contribution
~\cite{Yang:2026commu}, the resulting distributions reproduce the
overall behavior of the BESIII data.

\section{Conclusions}\label{sec4}

We have performed a theoretical study of the
$\eta_c\to p\bar p\eta$ decay, which has recently been observed by the
BESIII Collaboration through the radiative transition
$\psi(3686)\to\gamma\eta_c$~\cite{BESIII:2026kdm}, by combining the
SU(3) flavor structure of the primary production process with the
final-state interactions in the $p\eta$ and $\bar p\eta$ subsystems.
The primary source is described by two independent single-trace SU(3)
structures. These structures produce the $\pi N$, $\eta N$,
$K\Lambda$, and $K\Sigma$ channels with weights determined by the
parameters $\widetilde A$ and $\widetilde B$. The meson--baryon
scattering amplitudes are obtained by solving the coupled-channel
Bethe--Salpeter equation within the chiral unitary approach.

The low-energy $S_{11}$ dynamics is contained in the complete
coupled-channel scattering amplitude. In this framework, the prominent
$N^*(1535)$ structure is generated dynamically from the interaction
of the $\pi N$, $\eta N$, $K\Lambda$, and $K\Sigma$ channels. The
direct-production and rescattering mechanisms are combined coherently
in the total amplitude, and their interference is retained in the
invariant-mass distributions.

The three parameters $\widetilde A$, $\widetilde B$, and
$\widetilde C$ are determined simultaneously from the fit to the
$p\eta$ and $\bar p\eta$ invariant-mass distributions.
\textcolor{black}{Two solutions are obtained, with
$\chi^2/\mathrm{d.o.f.}=1.37$ and $1.39$, respectively.
The first solution has a relatively weak tree-level contribution,
whereas the second exhibits a much stronger tree-level contribution.
The latter is accompanied by substantial destructive interference
between the tree-level and rescattering amplitudes, as well as
between the tree-level and $N^*(1895)$ amplitudes.
The interference between the rescattering and $N^*(1895)$
amplitudes provides a constructive contribution in the
intermediate-mass region.
The two solutions therefore reproduce the fitted mass distributions
with different combinations of the production mechanisms.}
The parameter $\widetilde C$ determines the two charge-conjugate
$N^*(1895)$ contributions, which are included coherently and
contribute mainly to the higher invariant-mass region.

The $p\eta$ and $\bar p\eta$ distributions are described by the same
charge-conjugate dynamics and provide the main constraints on the
three fit parameters. The $p\bar p$ spectrum is obtained from the
same three-body amplitude. The broad structure in the $p\bar p$
spectrum is interpreted as a kinematic reflection of the
meson--nucleon dynamics, without introducing an independent
$p\bar p$ interaction.
\textcolor{black}{The first solution gives a better description of the
$p\bar p$ mass distribution and is therefore favored by this
additional comparison.}

The present results demonstrate the importance of meson--baryon
final-state interactions in the $\eta_c\to p\bar p\eta$ decay. The
results are consistent with a description in which the $N^*(1535)$
is dynamically generated by meson--baryon interactions. At the same
time, the fitted higher-mass contributions indicate that additional
$S_{11}$ strength may be relevant near the $N^*(1895)$ region.
\textcolor{black}{The presence of two fitted solutions illustrates the
role of interference among the different production mechanisms.}
More precise measurements of the three invariant-mass distributions
would be useful for reducing the correlations among the fitted
parameters and for clarifying the origin of the higher-mass
contribution.

\section*{Acknowledgments}
This work is partly supported by the
National Natural Science Foundation of China (NSFC)
under Grants No. 12405089, and by the China Postdoctoral Science Foundation under Grant No. 2022M720360
and No. 2022M720359. J. Song would also like to thank the support from the Hainan Provincial Excellent Talent Team under the “Four Talents” Gathering Program of Hainan Province.
Yu-Shan Ren acknowledges the support from the China Scholarship Council.
This work is also supported by the Grants PID2023-
147458NB-C21 and CEX2023-001292-S funded by MI-
CIU/AEI/10.13039/501100011033 and by ERDF/EU,
as well as of the PROMETEO program Grant
CIPROM/2023/59 funded by Generalitat Valenciana
10.13039/501100003359.

\bibliography{refs.bib} 
\end{document}